%% file: fracocc.tex
\documentclass[english,aps,pra,twocolumn,longbibliography]{revtex4-1}
\usepackage[T1]{fontenc}
\usepackage[latin9]{inputenc}
\usepackage{refstyle}
\usepackage{amsmath}

\makeatletter

\AtBeginDocument{\providecommand\subsecref[1]{\ref{subsec:#1}}}
\AtBeginDocument{\providecommand\eqref[1]{\ref{eq:#1}}}
\AtBeginDocument{\providecommand\tabref[1]{\ref{tab:#1}}}
\providecommand{\tabularnewline}{\\}
\RS@ifundefined{subsecref}
  {\newref{subsec}{name = \RSsectxt}}
  {}
\RS@ifundefined{thmref}
  {\def\RSthmtxt{theorem~}\newref{thm}{name = \RSthmtxt}}
  {}
\RS@ifundefined{lemref}
  {\def\RSlemtxt{lemma~}\newref{lem}{name = \RSlemtxt}}
  {}

\usepackage{braket}
\usepackage[version=3]{mhchem}
\usepackage{url}
\usepackage[shortcuts]{extdash}

\newref{sec}{name = section~, names = sections~, Name = Section~, Names = Sections~}
\newref{subsec}{name = section~, names = sections~}

\usepackage{subfig}
\usepackage{notes2bib}

\makeatother

\usepackage{babel}
\begin{document}
\title{Fully numerical calculations on atoms with fractional occupations.
Range-separated exchange functionals}
\author{Susi Lehtola}
\address{Department of Chemistry, University of Helsinki, P.O. Box 55 (A. I. Virtasen
aukio 1), FI-00014 Helsinki, Finland.}
\email{susi.lehtola@alumni.helsinki.fi}

\selectlanguage{english}%
\begin{abstract}
A recently developed finite element approach for fully numerical atomic
structure calculations {[}S. Lehtola, Int. J. Quantum Chem. 119, e25945
(2019){]} is extended to the description of atoms with spherically
symmetric densities via fractionally occupied orbitals. Specialized
versions of Hartree--Fock as well as local density and generalized
gradient approximation density functionals are developed, allowing
extremely rapid calculations at the basis set limit on the ground
and low-lying excited states even for heavy atoms. 

The implementation of range-separation based on the Yukawa or complementary
error function (erfc) kernels is also described, allowing complete
basis set benchmarks of modern range-separated hybrid functionals
with either integer or fractional occupation numbers. Finally, computation
of atomic effective potentials at the local density or generalized
gradient approximation levels for the superposition of atomic potentials
(SAP) approach {[}S. Lehtola, J. Chem. Theory Comput. 15, 1593 (2019){]}
that has been shown to be a simple and efficient way to initialize
electronic structure calculations is described. 

The present numerical approach is shown to afford beyond microhartree
accuracy with a small number of numerical basis functions, and to
reproduce literature results for the ground states of atoms and their
cations for $1\leq Z\leq86$. Our results indicate that the literature
values deviate by up to 10 $\mu E_{h}$ from the complete basis set
limit. The numerical scheme for the erfc kernel is shown to work by
comparison to results from large Gaussian basis set calculations from
the literature. Spin-restricted ground states are reported for Hartree--Fock
and Hartree--Fock--Slater calculations with fractional occupations
for $1\leq Z\leq118$.
\end{abstract}
\maketitle
\global\long\def\ERI#1#2{(#1|#2)}%
\global\long\def\bra#1{\Bra{#1}}%
\global\long\def\ket#1{\Ket{#1}}%
\global\long\def\braket#1{\Braket{#1}}%
\global\long\def\erf#1{\text{erf }#1}%
\global\long\def\erfc#1{\text{erfc }#1}%

\newcommand*\citeref[1]{ref. \citenum{#1}}
\newcommand*\citerefs[1]{refs. \citenum{#1}}
\newcommand\HelFEM{\textsc{HelFEM}}
\newcommand\Erkale{\textsc{Erkale}}
\newcommand\libxc{\textsc{libxc}}\newcommand\PsiFour{\textsc{Psi4}}
\newcommand\xtwodhf{\textsc{x2dhf}}
\newcommand\apcinfty{\mbox{aug-pc-$\infty$}}

\section{Introduction\label{sec:Introduction}}

Atoms are the simplest possible unit in chemistry, which is why electronic
structure studies on atoms have a long and venerated history. Thanks
to the high amount of symmetry that may be used to reduce the number
of degrees of freedom in the atomic problem, fully numerical electronic
structure approaches on atoms have been possible for a very long time
\citep{Lehtola2019c}; for instance, a fully numerical configuration
interaction calculation on the oxygen atom was reported by Hartree
and coworkers over 80 years ago \citep{Hartree1939}.

As the atomic hamiltonian is spherically symmetric, the exact wave
function should be rotationally invariant as well. Although the necessary
symmetry requirements can straightforwardly be enforced in wave function
approaches, the application of density functional theory \citep{Hohenberg1964,Kohn1965}
(DFT) on atoms is surprisingly tricky. In the usual DFT approach,
a single Slater determinant is employed, with all orbitals below the
Fermi level being fully occupied. Non-relativistically, all $2l+1$
atomic orbitals sharing the principal quantum number $n$ and angular
quantum number $l$ should be completely degenerate; however, this
behavior is broken by conventional DFT as well as Hartree--Fock (HF)
already on the first row. Different choices for the occupied orbitals
on the $2p$ shell yield different final energies for e.g. B and F,
which may lead to several kcal/mol differences in the total energy
-- with a symmetrized density yielding yet another result \citep{Baerends1997a}.
One possibility to obtain comparable results is to employ a standard
set of electronic occupations \citep{Hay1977}, but such an approach
does not yield the lowest possible energy. 

Pursuing the lowest energy is not unproblematic, either. While HF
is infamous for possessing variational solutions that break symmetry
in systems with a high degree of symmetry \citep{Prat1972}, symmetry
breaking is a problem in DFT as well \citep{Goursot1995}. In atoms,
broken symmetries often arise for open shells, and the effect of non-spherical
densities is known to be more pronounced with functionals at the generalized
gradient approximation (GGA) and especially the meta-GGA (mGGA) level
than at the local density approximation (LDA) level \citep{Kutzler1987,Philipsen1996,Tao2005,Johnson2007};
even optimized effective potential exact-exchange calculations are
subject to spurious energy splittings \citep{Pittalis2006}. Inclusion
of current density dependence leads to improvement of GGA and mGGA
results \citep{Tao2005,Johnson2007}, but the proper orbital degeneracy
is still not fully restored. 

Symmetry breaking effects in atoms can be seen already at the simplest
possible level of DFT, that is, the exchange-only LDA, which is also
commonly known as Hartree--Fock--Slater (HFS) theory. For example,
HFS calculations on the F atom reveal millihartree decreases of the
total energy upon addition of $d$ as well as $f$ functions, which
is at variance to the generally accepted electronic configuration
of fluorine as $1s^{2}2s^{2}2p^{5}$. Interestingly, this kind of
symmetry breaking sometimes happens even in the case of closed-shell
atoms; see, for instance, our recent finite element reproduction \citep{Lehtola2019a}
of calculations on atomic anions \citep{Anderson2017} where symmetry
breaking was observed for \ce{H-}, Be, \ce{Li-}, and \ce{Na-}. 

In addition to being degenerate due to symmetry (as often in atoms),
orbitals may also be degenerate by accident. Since the aufbau rule
implies populating the orbitals in increasing energy, it tempting
to divide the occupations evenly in the case of degeneracies. This
paves the way to the use of fractional occupations, which in the case
of atoms naturally yield a spherically symmetric density thanks to
Unsöld's theorem \citep{Unsold1927}; the use of fractional occupations
can be formally justified within the theory of ensemble representable
densities \citep{Englisch1984,Englisch1984a}. 

Fractionally occupied orbitals should especially be used in the case
where there is a negative gap between the highest occupied and lowest
unoccupied orbital no matter which way the orbitals are occupied;
this happens when the highest occupied and lowest unoccupied orbital
switch places during the orbital optimization. In this case, the total
energy can be lowered by moving a fraction of an electron from the
highest occupied orbital to the lowest unoccupied orbital, and at
some point the two levels should cross. 

Fractional occupations have been shown to yield better results for
strongly correlated systems \citep{Dunlap1983a,Wang1996,Schipper1998,Takeda2003,Nygaard2013a}.
However, fractional occupations can only be justified at the Fermi
level \citep{Valiev1995}, and more recently it has been shown that
energy minimization naturally leads to integer occupations below the
Fermi level, and possible fractional occupations at the Fermi level
for independent particle models like HF and DFT \citep{Giesbertz2010}.

While in some systems it is clear \emph{a priori} from symmetry arguments
or the orbital energies how many orbitals should be fractionally occupied,
this is generally not the case. However, fractional occupations can
be obtained as \citep{Kraisler2009} the zero-temperature limit of
finite-temperature DFT (FT-DFT) \citep{Mermin1965,Stoitsov1988}.
In a finite-temperature approach, the fractional orbital occupation
numbers are determined by the orbital energies according to some smearing
scheme that is typically controlled by a single parameter, an electronic
temperature. Because of the simplicity and favorable computational
scaling of FT-DFT, it has become a powerful tool for approximate modeling
of systems exhibiting strong correlation; such approaches have been
used to obtain promising results for a variety of systems \citep{Chai2012,Chai2014,Wu2015a,Yeh2016,Seenithurai2016,Wu2016,Chai2017,Seenithurai2017,Lin2017,Yeh2018,Seenithurai2018,Deng2019,Chung2019,Grimme2013a,Grimme2015}.

Finite electronic temperatures may also be used to aid the convergence
of self-consistent field calculations of molecules \citep{Rabuck1999};
in the solid state, the use of fractional occupation numbers is often
mandatory in order to attain convergence \citep{Kratzer2019}. Although
finite temperature approaches are more attractive for DFT where all
electrons experience the same potentials, finite temperature approaches
can also be used in the context of HF calculations where they may
offer good active spaces for post-HF calculations on strongly correlated
systems \citep{Slavicek2010}.

Although several types of smearing schemes have been suggested, including
Fermi--Dirac \citep{Mermin1965}, Gaussian smearing \citep{Fu1983},
Methfessel--Paxton smearing \citep{Methfessel1989}, cold smearing
\citep{Marzari1999}, and others \citep{Holender1995}, they have
been shown to yield similar results if the parameters are adjusted
properly \citep{Springborg1998,Grotheer1998,Slavicek2010}; however,
the behavior with respect to temperature needs to be carefully checked
in each case to ensure convergence \citep{Mehl2000}. Note that the
evaluation of forces in finite-temperature calculations require the
consideration of an additional entropic term that arises from the
non-integer occupations and that depends on the smearing function
\citep{Weinert1992,Warren1996}.

Regardless of the used temperature, calculations with fractional occupations
are more involved than those with integer occupations. Convergence
acceleration techniques such as direct inversion in the iterative
subspace \citep{Pulay1980,Pulay1982} (DIIS) become invalid when the
orbital occupation pattern changes, even though the self-consistent
field problem itself may become easier with fractional occupation
numbers \citep{Rabuck1999}. Determining the correct occupations is
hard, since the orbital occupations depend on the orbital energies,
which in turn depend on the orbital occupations. The changes in the
occupations may also cause changes in the shapes of the orbitals,
meaning that the orbitals, their energies and their occupations need
to be solved self-consistently. Several approaches have been proposed
for solving this problem both for zero \citep{Averill1992,Cances2003,Kraisler2009,Nygaard2013}
and finite electronic temperatures \citep{Gillan1989,Fernando1989,Grumbach1994,Chetty1995,Holender1995,Marzari1997a}.

In systems with a high degree of symmetry such as atoms, the fractional
occupations can be defined by symmetry block. Fractional occupations
for atoms are typically defined in terms of atomic shells, over which
the electrons are equally divided. For instance, the $1s^{2}2s^{2}2p^{5}$
configuration for F implies that the hole in the $2p$ shell be equally
divided, resulting in the minority spin occupations $2p_{x}^{2/3}2p_{y}^{2/3}2p_{z}^{2/3}$;
a spin-restricted variant would employ occupations of $2p_{x}^{5/6}2p_{y}^{5/6}2p_{z}^{5/6}$
in both spin channels. Indeed, this is the method of choice for fully
numerical density functional calculations on atoms \citep{Lehtola2019c},
and it has been used \emph{e.g. }in \citeref{Kotochigova1997} for
local density calculations on $1\leq Z\leq92$ at the ground state
electronic configuration from experiment, and in \citeref{Lee1997a}
for Perdew--Burke--Ernzerhof (PBE) \citep{Perdew1996,Perdew1997}
calculations on $1\leq Z\leq20$ and $31\leq Z\leq36$. 

Atomic calculations with fractional occupation numbers are also typically
used to generate pseudopotentials \citep{Fuchs1999,Oliveira2008},
numerical atomic orbital basis functions \citep{Ozaki2004,Blum2009},
and Gaussian basis sets \citep{Andzelm1985,Godbout1992,Porezag1999}.
Spin-restricted spherically symmetric atoms may also be used for setting
up frozen core calculations within all-electron approaches, and to
determine approximate binding energies \citep{TeVelde2001}. We have
also recently shown that the radial potential from atomic calculations
with fractional occupation numbers can be used to formulate efficient
initial guesses for electronic structure calculations on polyatomic
systems via the superposition of atomic potentials (SAP) approach
\citep{Lehtola2019}.

In the typical case, electrons are divided evenly among the $2l+1$
orbitals that are degenerate by symmetry. However, the fractional
occupations can be generalized beyond integer occupations per shell,
in case accidental degeneracy is also present. Early multiconfigurational
HF calculations on atoms found that the $3d$ orbitals become occupied
before the $4s$ orbitals in transition metals \citep{Slater1969,Abdulnur1972},
which was solved by moving fractions of an electron between the shells.
One example of this approach is the iron atom, where the {[}Ar{]}$3d_{1}^{5}4s_{1}^{1}$
and {[}Ar{]}$3d_{2}^{5}4s_{0}^{1}$ configurations both turn out to
have a negative gap in the local-density approximation \citep{Janak1978},
the upper and lower indices denoting spin-up and spin-down electrons,
respectively. With the Vosko--Wilk--Nusair (VWN) local density functional,
the lowest-energy configuration is found to be {[}Ar{]}$3d_{1.3984}^{5}s_{0.602}^{1}$
\citep{Kraisler2009}. 

A systematic, non-relativistic study for spherical atoms $1\leq Z\leq86$
has recently been presented by Kraisler, Makov and Kelson for the
local density and PBE functionals based on three local density functionals,
employing 16~000 point grids and wave functions converged to 2 $\mu E_{h}$
\citep{Kraisler2010}. It was found in \citeref{Kraisler2010} that
the ground state of most atoms does not involve fractional splitting
of electrons between shells, indicating that a fully numerical program
for modeling atoms with spherical densities would go a long way towards
the final solution. 

While several programs exist for either wave function or density functional
based fully numerical calculations on atoms \citep{Lehtola2019c},
we are not aware of any publicly available software that supports
hybrid functionals, except the recently published \HelFEM{} program
\citep{Lehtola2019a,HelFEM}, which also includes a fully numerical
approach for diatomic molecules that similarly supports hybrid functionals
\citep{Lehtola2019b}. Most publicly available programs for fully
numerical density functional calculations on atoms target the generation
of projector-augmented wave (PAW) setups \citep{Blochl1994} or the
generation of pseudopotentials \citep{Schwerdtfeger2011}. Although
Hartree--Fock pseudopotential generators have been available for
some time \citep{Trail2005,Al-Saidi2008}, which allowed the use of
non-self-consistent pseudopotentials for hybrid functionals \citep{Wu2009b},
surprisingly, the self-consistent generation of pseudopotentials for
hybrid functionals has only been described last year \citep{Yang2018},
explaining the scarcity of such programs. 

Interestingly, the work of Yang et al in \citeref{Yang2018} did not
employ fractionally occupied Hartree--Fock calculations, but rather
followed Slater's multiconfigurational approach, which is at odds
with the density functional description used in the work, as the exact
exchange and density functional parts experience different electron
densities. In contrast, when fractional occupations are employed as
in the present work, the exchange exact operator becomes independent
of the magnetic quantum number $m$ as will be shown in \subsecref{Self-consistent-field-calculatio},
and both the density functional and exact exchange operators are evaluated
with the same density matrix.

Although a general-use atomic program like the one in \HelFEM{} can
be straightforwardly adapted to calculations on spherically symmetric
densities by employing fractional occupation numbers in the construction
of the density matrix, a more efficient approach is afforded by taking
the assumption of the spherical symmetry of the density matrix deeper
in the algorithms. As a result, some or even all of the angular integrals
can be eliminated from the calculations, reducing the problem to a
small number of dimensions; indeed, this is\emph{ }exactly what is
done in the multiconfigurational HF approach Slater proposed 90 years
ago \citep{Slater1929}.

In the present work, we describe the extension of the atomic program
in \HelFEM{} to the description of atoms with spherical symmetric
density via fractional occupation numbers. Alike the other programs
in \HelFEM{}, the spherically symmetric atomic program is interfaced
to the \libxc{} library of density functionals \citep{Lehtola2018}
and can be used with all supported density functionals therein. Specialized
implementations for atomic calculations with fractional occupations
are developed for local density (LDA) and generalized gradient (GGA)
functionals as well as HF exchange, yielding significant reductions
in the dimensionality of the problem, whereas meta-GGA functionals
can be used via an interface to the algorithms previously developed
in \citeref{Lehtola2019a}. 

Importantly, we also describe the implementation of Yukawa and complementary
error function (erfc) range-separated exchange for atomic calculations
in \HelFEM{} with either fractional or integer occupations, allowing
complete basis set benchmarks of recently developed exchange-correlation
functionals such as the CAM-QTP family by Bartlett and coworkers \citep{Verma2014,Jin2016,Haiduke2018},
the N12-SX and revM11 functionals by Truhlar and coworkers \citep{Peverati2012a,Verma2019},
and the $\omega$B97X-V and $\omega$B97M-V functionals by Mardirossian
and Head-Gordon (without the non-local correlation part) \citep{Mardirossian2014a,Mardirossian2016}.
While the spherical harmonics decomposition for the Yukawa kernel
is well known, the decomposition for the erfc kernel has only been
derived some time ago \citep{Angyan2006} and does not appear to have
been implemented within a generally applicable fully numerical approach
for atoms. Results for H and He with relatively low-order B-spline
basis sets have, however, been published almost simultaneously to
our work \citep{Zapata2019}. Finally, we also describe the analytic
calculation of the radial potentials necessary for the SAP orbital
guess \citep{Lehtola2019}.

In the next section, we derive the equations for fractionally occupied
HF and DFT at the LDA and GGA levels within the used finite element
approach. Then, in the Results section, we present applications of
the program to reproducing ground states for the neutral atoms and
cations $1\leq Z\leq86$ and compare with \citeref{Kraisler2010};
we reproduce the long-range corrected density functional calculations
on closed-shell atoms of \citeref{Anderson2017} to show that the
range-separation scheme works; and finally we report the non-relativistic
ground states of all atoms in the periodic table at HF and HFS levels
of theory. The article concludes with a brief summary and discussion
section.

\section{Method\label{sec:Method}}

A basis set of the form 
\begin{equation}
\chi_{nlm}=r^{-1}B_{n}(r)Y_{l}^{m}(\theta,\phi)\label{eq:basis}
\end{equation}
is adopted as in the integer-occupation program described in \citeref{Lehtola2019a}.
Here, $B_{n}(r)$ are the piecewise polynomial shape functions of
the finite element method, which have been discussed extensively in
\citerefs{Lehtola2019c} and \citenum{Lehtola2019a} to which we refer
for further details.

\subsection{Range-separated exchange\label{subsec:Range-separated-exchange}}

As discussed in \citerefs{Lehtola2019c} and \citenum{Lehtola2019a},
the key to fully numerical electronic structure calculations on atoms
is the Laplace expansion
\begin{align}
\frac{1}{r_{12}}= & \frac{4\pi}{r_{>}}\sum_{L=0}^{\infty}\frac{1}{2L+1}\left(\frac{r_{<}}{r_{>}}\right)^{L}\sum_{M=-L}^{L}Y_{L}^{M}(\Omega_{1})\left(Y_{L}^{M}(\Omega_{2})\right)^{*}\label{eq:laplace}
\end{align}
that factorizes the two-electron integrals
\begin{align}
(ij|kl)= & \int\frac{\chi_{i}(\boldsymbol{r})\chi_{j}^{*}(\boldsymbol{r})\chi_{k}(\boldsymbol{r}')\chi_{l}^{*}(\boldsymbol{r}')}{\left|\boldsymbol{r}-\boldsymbol{r}'\right|}{\rm d}^{3}r{\rm d}^{3}r'\label{eq:tei0}
\end{align}
into a radial and an angular part. 

In range-separated density functional theory \citep{Gill1996c,Leininger1997},
the Coulomb interaction is split into a short-range (sr) and a long-range
(lr) part as\begin{widetext}
\begin{equation}
\frac{1}{r}=\frac{\phi_{\text{sr}}(r)}{r}+\frac{1-\phi_{\text{sr}}(r)}{r}=\frac{\phi_{\text{sr}}(r)}{r}+\frac{\phi_{\text{lr}}(r)}{r}=\frac{1-\phi_{\text{lr}}(r)}{r}+\frac{\phi_{\text{lr}}(r)}{r},\label{eq:rangesep}
\end{equation}
\end{widetext} where $\phi_{{\rm \text{sr}}}(r)=1-\phi_{\text{lr}}(r)$
is a splitting function. Typically, the short-range part is described
using density functional theory, and the long-range part with HF theory,
but in practice many functionals employ more flexibility: for instance,
the CAM-B3LYP functional \citep{Yanai2004b} contains 19\% short-range
and 65\% long-range exact exchange. 

The evaluation of the range-separated exchange functionals is simple
if one has access to the Green's function expansion of the range-separated
kernel as
\begin{align}
\frac{\phi_{{\rm sr}}(r)}{r_{12}} & =\sum_{L=0}^{\infty}\frac{4\pi}{2L+1}\mathcal{G}_{L}(r_{<},r_{>},\mu)\nonumber \\
 & \times\sum_{M=-L}^{L}\left(Y_{L}^{M}(\Omega_{1})\right)^{*}Y_{L}^{M}(\Omega_{2})\label{eq:green}
\end{align}
where $\mathcal{G}_{L}(r_{>},r_{<},\mu)$ is the Green's function,
$r_{>}$ and $r_{<}$ are the greater and smaller of $r_{1}$ and
$r_{2}$, respectively, and $\mu$ is the range separation parameter.
The Green's function for the (unscreened) classical Coulomb interaction
can be identified from \eqref{laplace} as
\begin{equation}
\mathcal{G}_{L}^{\text{Coulomb}}(r_{>},r_{<})=\frac{r_{<}^{L}}{r_{>}^{L+1}}.\label{eq:Gl-coulomb}
\end{equation}
The implementation of the integrals in \HelFEM{} is based on the
primitive integrals defined in \citeref{Lehtola2019a} as
\begin{align}
I_{ijkl}^{L} & =\frac{4\pi}{2L+1}\int{\rm d}r_{1}{\rm d}r_{2}B_{i}(r_{1})B_{j}(r_{1})\nonumber \\
 & \times B_{k}(r_{2})B_{l}(r_{2})\mathcal{G}_{L}(r_{>},r_{<},\mu),\label{eq:primint}
\end{align}
where $B_{i}(r)$ are the piecewise polynomial basis functions of
\eqref{basis}.

\subsubsection{Yukawa kernel\label{subsec:Yukawa-kernel}}

The Yukawa-screened \citep{Yukawa1935} potential, $\phi^{\text{sr}}(r_{12})=\exp(-\lambda r_{12})$
has a relatively well-known simple expansion
\begin{align}
\frac{e^{-\lambda|\boldsymbol{r}-\boldsymbol{r}'|}}{|\boldsymbol{r}-\boldsymbol{r}'|} & =4\pi\lambda\sum_{L=0}^{\infty}i_{L}(\lambda r_{<})k_{L}(\lambda r_{>})\nonumber \\
 & \times\sum_{M=-L}^{L}Y_{L}^{M}(\Omega_{1})\left(Y_{L}^{M}(\Omega_{2})\right)^{*}\label{eq:sr-yukawa}
\end{align}
 where $i_{L}$ and $k_{L}$ are regular and irregular modified Bessel
functions that are regular at zero and infinity, respectively. Due
to its separability, Yukawa-screened functionals are easy to handle
in fully numerical approaches. Indeed, the Yukawa Green's function
is employed in several recently developed linear scaling approaches
for solving the HF or Kohn--Sham equations for bound orbitals in
molecular systems via the Helmholtz kernel \citep{Harrison2004,Frediani2013,Jensen2014,Solala2017,Parkkinen2017}.
The Yukawa interaction is also straightforward to implement in calculations
with Slater-type orbitals \Citep{Seth2012,Seth2013,Rico2013}. It
turns out that Yukawa screening can also be implemented with Gaussian-type
orbitals in a rather straightforward manner \citep{Akinaga2008},
as analogous integrals also arise within $r_{12}$ wave function theory
\citep{Ten-no2004,Ten-No2007}. Such implementations are, however,
rare at the moment, even though it has been claimed that Yukawa screening
yields more accurate atomization and charge transfer excitation energies
than erfc screening \citep{Akinaga2009}. The Green's function for
the Yukawa interaction can be read from \eqref{sr-yukawa} as
\begin{equation}
\mathcal{G}_{L}^{\text{Yukawa}}(r_{>},r_{<},\lambda)=(2L+1)\lambda i_{L}(\lambda r_{<})k_{L}(\lambda r_{>}).\label{eq:Gl-Yukawa}
\end{equation}
As the Yukawa interaction factorizes in $r_{>}$ and $r_{<}$, it
can be implemented in a similar fashion to the full Coulomb interaction,
\eqref{Gl-coulomb}, along the lines of \citeref{Lehtola2019a}.

\subsubsection{erfc kernel\label{subsec:erfc-kernel}}

Most range-separated functionals, however, are based on the complementary
error function (erfc) kernel $\phi_{\text{sr}}(r)=\erfc{(\mu r)}$.
Such functionals are easy to implement in Gaussian-basis programs,
requiring but simple modifications to the two-electron integrals \citep{Heyd2003,Ahlrichs2006},
as well as plane wave programs since the kernel has a simple Fourier
transform which is strongly attenuated at large momentum. In contrast,
the implementation of the erfc kernel is more complicated in real-space
approaches. Fortunately, spherical harmonic expansions for the erfc
Green's functions are available in the literature \citep{Marshall2002,Angyan2006},
but their form is more involved than that of the Yukawa function in
\eqref{sr-yukawa}. The main complication is that the Green's function
does not factorize in $r_{<}$ and $r_{>}$, which means that two-dimensional
quadrature is always required. In the approach of \citeref{Angyan2006},
new variables are introduced as $\Xi=\mu R$ and $\xi=\mu r$ and
\begin{equation}
\mathcal{G}_{L}(R,r;\mu)=\mu\Phi_{L}(\Xi,\xi)\label{eq:Gl-erfc}
\end{equation}
where $\Phi_{L}$ is a scaled radial function given by \begin{widetext}
\begin{align}
\Phi_{n}(\Xi,\xi) & =F_{n}(\Xi,\xi)+\sum_{m=1}^{n}F_{n-m}(\Xi,\xi)\frac{\Xi^{2m}+\xi^{2m}}{\left(\Xi\xi\right)^{m}}H_{n}(\Xi,\xi)\label{eq:phin}\\
F_{n}(\Xi,\xi) & =\frac{2}{\sqrt{\pi}}\sum_{p=0}^{n}\left(-\frac{1}{4\Xi\xi}\right)^{p+1}\frac{(n+p)!}{p!(n-p)!}\left[(-1)^{n-p}e^{-\left(\xi+\Xi\right)^{2}}-e^{-\left(\xi-\Xi\right)^{2}}\right]\label{eq:Fn}\\
H_{n}(\Xi,\xi) & =\frac{1}{2(\xi\Xi)^{n+1}}\left[(\Xi^{2n+1}+\xi^{2n+1})\erfc{(\Xi+\xi)-(\Xi^{2n+1}-\xi^{2n+1})\erfc{(\Xi-\xi)}}\right]\label{eq:Hn}
\end{align}
\end{widetext} (Note that the lower limit of the sum in \eqref{Fn}
is incorrect in \citeref{Angyan2006}, where it reads $p=1$ instead
of $p=0$.) \Eqrangeref{phin}{Hn} are numerically unstable in the
short range, which is why when either $\xi<0.4$, or $\Xi<0.5$ and
$0<\xi<2\Xi$ \citep{Angyan2006}, the Green's function is evaluated
with a Taylor expansion\begin{widetext}
\begin{align}
\Phi_{n}(\Xi,\xi) & =\sum_{k}\frac{D_{n,k}(\Xi)}{\Xi^{n+1}}\xi^{n+2k},\label{eq:Phin-short}\\
D_{n,0}(\Xi) & =\erfc{\Xi}+\frac{\exp(-\Xi^{2})}{\sqrt{\pi}}\left(2\Xi^{2}\right)^{n+1}\sum_{m=1}^{n}\frac{\left(2\Xi^{2}\right)^{-m}}{(2n-2m-1)!!}\label{eq:Dn0}\\
D_{n,k}(\Xi) & =\frac{\exp(-\Xi^{2})}{\sqrt{\pi}}\left(2\Xi^{2}\right)^{n+1}\frac{2n+1}{k!(2n+2k+1)}\sum_{m=1}^{k}{m-k-1 \choose m-1}\frac{\left(2\Xi^{2}\right)^{k-m}}{(2n+2k-2m-1)!!},\ k\geq0\label{eq:Dnk}
\end{align}
\end{widetext}

Despite the lack of factorization of the erfc Green's function, its
evaluation can be carried out analogously to the Coulomb and Yukawa
kernels. The primitive integrals, \eqref{primint}, can be divided
into two cases thanks to the finite support of the piecewise polynomial
basis functions, as discussed in \citeref{Lehtola2019a}. In an intraelement
integral, both $ij$ and $kl$ are within the same element, whereas
in an interelement integral $ij$ are in one element and $kl$ are
in another. In analogy to the scheme for Coulomb integrals discussed
in \citeref{Lehtola2019a}, the interelement integrals are evaluated
with $N_{\text{quad}}$ quadrature points in both $ij$ and $kl$,
whereas the intraelement integrals employ $N_{\text{quad}}$ points
in $ij$, whereas the $kl$ quadrature is split into $N_{\text{quad}}$
intervals, all of which employ a fresh set of $N_{\text{quad}}$ quadrature
points.

\subsection{Self-consistent field calculations with fractional occupations\label{subsec:Self-consistent-field-calculatio}}

It is well known that atomic orbitals can be written in the form
\begin{equation}
\psi_{nlm}(\boldsymbol{r})=R_{nl}(r)Y_{l}^{m}(\hat{\boldsymbol{r}}).\label{eq:atorb}
\end{equation}
Employing smeared occupations as
\begin{align}
n(\boldsymbol{r}) & =\sum_{n=1}^{\infty}\sum_{l=n-1}^{\infty}f_{nl}\sum_{m=-l}^{l}|\psi_{nlm}(\boldsymbol{r})|^{2}\label{eq:n-smear}\\
 & =\sum_{nl}f_{nl}\frac{\left(2l+1\right)R_{nl}^{2}(r)}{4\pi}=n(r)\label{eq:n-spherical}
\end{align}
where $f_{nl}$ is the occupation number of all the $2l+1$ orbitals
on the $(n,l)$ shell, one immediately sees that the density matrix
is diagonal in $l$ and $m$
\begin{equation}
P_{\mu\nu}^{\sigma}=\delta_{l_{\mu},l_{\nu}}\delta_{m_{\mu},m_{\nu}}P_{\mu\nu}^{l_{\mu};\sigma}\label{eq:densmat}
\end{equation}
and that the elements of the density matrix only depend on the value
of $l$.

The spherical averaging yields huge simplifications for density functional
calculations. As now the density is only a function of the radial
coordinate, also its gradient
\begin{equation}
\nabla n=\partial_{r}n\hat{\boldsymbol{e}}_{r}\label{eq:gradn}
\end{equation}
only depends on the radial coordinate. Following the usual projective
approach \citep{Pople1992,Lehtola2019a}, the LDA and GGA matrix elements
\begin{widetext}
\begin{align}
K_{\mu\nu}^{xc;\sigma}= & \int\Bigg[\frac{\delta f_{\text{xc}}}{\delta n_{\sigma}\left(\mathbf{r}\right)}\phi_{\mu}(\mathbf{r})\phi_{\nu}(\mathbf{r})+\Bigg(2\frac{\delta f_{\text{xc}}}{\delta\gamma_{\sigma\sigma}\left(\mathbf{r}\right)}\nabla\rho_{\sigma}(\mathbf{r})+\frac{\delta f_{\text{xc}}}{\delta\gamma_{\sigma\sigma'}\left(\mathbf{r}\right)}\nabla\rho_{\sigma'}(\mathbf{r})\Bigg)\cdot\nabla\left(\phi_{\mu}(\mathbf{r})\phi_{\nu}(\mathbf{r})\right)\Bigg]{\rm d}^{3}r\label{eq:Kxc}
\end{align}
\end{widetext} become greatly simplified as only the radial terms
are picked up, and as the same radial basis is used for all $l,m$;
see \eqref{basis}. Note, however, that meta-GGAs that depend on the
kinetic energy density cannot be handled in the same fashion, as the
kinetic energy density is not manifestly dependent only on the radial
coordinate as discussed e.g. in \citeref{DellaSala2016}. Alike the
exact exchange discussed below, the meta-GGA potential turns out to
depend on the $l$ channel. Meta-GGA functionals can be used in the
present program via a fractional-occupation interface to the full
atomic routines discussed in \citeref{Lehtola2019a}.

The Coulomb matrix arising from \eqref{laplace} trivially reduces
to a single term as the spherically symmetric density only consists
of a single $L=0$, $M=0$ component. Exact exchange -- either with
the full Coulomb form of \eqref{Gl-coulomb} or the range-separated
versions in \eqref{Gl-Yukawa, Gl-erfc} -- is a bit more complicated,
as both the integrals and the density matrix carry a dependence on
the orbital angular momenta in the well-known equation
\begin{equation}
K_{\mu\nu}=\sum_{\sigma\tau}(\mu\sigma|\nu\tau)P_{\sigma\tau}.\label{eq:K}
\end{equation}
Employing the blocking of the density matrix given in \eqref{densmat},
the exchange matrix can be written as\begin{widetext}
\begin{align}
K_{\mu\nu}^{l_{\text{out}}} & =\sum_{\sigma\tau}(\mu\sigma|\nu\tau)P_{\sigma\tau}=\sum_{L=\left|l_{\text{in}}-l_{\text{out}}\right|}^{l_{\text{in}}+l_{\text{out}}}I_{\mu\sigma\nu\tau}^{L}P_{\sigma\tau}^{l_{\text{in}}}\frac{1}{2l_{\text{out}}+1}\sum_{m_{\text{in}}=-l_{\text{in}}}^{l_{\text{in}}}\sum_{m_{\text{out}}=-l_{\text{out}}}^{l_{\text{out}}}G_{Ll_{\text{in}},m_{\text{out}}}^{Mm_{\text{in}},l_{\text{out}}}G_{Ll_{\text{in}},m_{\text{out}}}^{Mm_{\text{in}},l_{\text{out}}}\label{eq:K-lout}
\end{align}
\end{widetext} where $L$ is a coupled angular momentum with $z$
projection $M=m_{\text{out}}-m_{\text{in}}$. Rearranging the contractions,
it is then seen that \begin{widetext} 
\begin{align}
K_{\mu\nu}^{l_{\text{out}}} & =\sum_{L}I_{\mu\sigma\nu\tau}^{L}\left(\sum_{l_{\text{in}}}P_{\sigma\tau}^{l_{\text{in}}}\left[\frac{1}{2l_{\text{out}}+1}\sum_{m_{\text{in}}=-l_{\text{in}}}^{l_{\text{in}}}\sum_{m_{\text{out}}=-l_{\text{out}}}^{l_{\text{out}}}G_{Ll_{\text{in}},m_{\text{out}}}^{Mm_{\text{in}},l_{\text{out}}}G_{Ll_{\text{in}},m_{\text{out}}}^{Mm_{\text{in}},l_{\text{out}}}\right]\right)\label{eq:K-rearr}
\end{align}
\end{widetext} where the evaluation is done from the insidemost bracket
out.

\subsection{Cusp condition\label{subsec:Cusp-condition}}

One way to diagnose atomic wave functions is the Kato--Steiner cusp
condition \citep{Kato1957,Steiner1963}
\begin{equation}
C=-\frac{1}{2Z}\frac{{\rm d}\log n(r)}{{\rm d}r}\Bigg|_{r=0}=-\frac{1}{2Z}\frac{n'(0)}{n(0)}\label{eq:cusp}
\end{equation}
which yields the value $C=1$ for the exact HF or density functional
solution \citep{Nagy2001}. The electron density $n(r)$ at the nucleus
was obtained in \citeref{Lehtola2019a} via l'Hôpital's rule as
\begin{align}
n(0) & =P_{\mu\nu}\lim_{r\to0}\frac{B_{\mu}(r)B_{\nu}(r)}{r^{2}}\label{eq:nucdens-def}\\
 & =P_{\mu\nu}\lim_{r\to0}\frac{\frac{{\rm d}^{2}}{{\rm dr}^{2}}B_{\mu}(r)B_{\nu}(r)}{\frac{{\rm d}^{2}}{{\rm dr}^{2}}r^{2}}\label{eq:nucdens-mid}\\
 & =P_{\mu\nu}B_{\mu}'(r)B_{\nu}'(r).\label{eq:nucdens}
\end{align}
Its derivative at the nucleus also turns out to have a simple expression:
\begin{widetext}
\begin{align}
n'(0) & =P_{\mu\nu}\lim_{r\to0}\left[\frac{B_{\mu}'(r)B_{\nu}(r)}{r^{2}}+\frac{B_{\mu}(r)B_{\nu}'(r)}{r^{2}}-2\frac{B_{\mu}(r)B_{\nu}(r)}{r^{3}}\right]\nonumber \\
 & =P_{\mu\nu}\lim_{r\to0}\left[\frac{\frac{{\rm d}^{3}}{{\rm dr}^{3}}B_{\mu}(r)B_{\nu}(r)}{\frac{{\rm d}^{2}}{{\rm dr}^{2}}r^{2}}-2\frac{\frac{{\rm d}^{3}}{{\rm dr}^{3}}B_{\mu}(r)B_{\nu}(r)}{\frac{{\rm d}^{3}}{{\rm dr}^{3}}r^{3}}\right]=P_{\mu\nu}B_{\mu}''(0)B_{\nu}'(0).\label{eq:nucder}
\end{align}
\end{widetext} The value of the cusp is printed out at the end of
all atomic calculations in \HelFEM{}.

\subsection{Effective radial potential for SAP\label{subsec:Effective-radial-potential}}

In the SAP approach discussed in \citeref{Lehtola2019}, approximate
orbitals for a molecule are obtained by diagonalizing an effective
one-body Hamiltonian in an external potential obtained as a superposition
of radial atomic potentials. Once the atomic ground state has been
found with any supported method in \HelFEM{}, including HF and hybrid
and meta-GGA functionals, the radial effective potential for the SAP
approach can be calculated based on any LDA or GGA functional. Extensions
to the exact exchange, as in the optimized effective potential method
\citep{Sharp1953}, as well as generalized Kohn--Sham methods for
the radial potentials from meta-GGA functionals are left for future
work. 

If the radial potential is self-consistent, i.e. the same functional
was used for both the atomic orbitals and the potential, the SAP guess
will reproduce the atomic orbitals exactly \citep{Lehtola2019}. The
atomic potential comprises Coulomb and exchange-correlation contributions,
the calculation of which is presented in the following.

\subsubsection{Coulomb potential\label{subsec:Coulomb-potential}}

Employing the Laplace expansion, \eqref{laplace}, the Coulomb potential
at a point\textbf{ $\boldsymbol{r}$} for a spherically symmetric
charge distribution is
\begin{align}
V(\boldsymbol{r}) & =\int_{0}^{\infty}\frac{1}{r_{>}}n(r')r^{2}{\rm d}r'\label{eq:Vsph}
\end{align}
Expressing the orbitals as in \eqref{atorb} yields potential matrix
elements of the form
\begin{align}
V_{ij}(r) & =\int_{0}^{\infty}\frac{1}{r_{>}}B_{i}(r')B_{j}(r'){\rm d}r'\label{eq:Vij}\\
 & =\frac{1}{r}\int_{0}^{r}B_{i}(r')B_{j}(r'){\rm d}r'\nonumber \\
 & +\int_{r}^{\infty}\frac{1}{r'}B_{i}(r')B_{j}(r'){\rm d}r'\label{eq:Vij-2}
\end{align}
 and one gets three cases depending on whether $r$ is inside the
element where $i$ and $j$ reside, or not. Let the element begin
at $r_{b}$ and end at $r_{e}$. Now \begin{widetext}
\begin{equation}
V_{ij}(r)=\begin{cases}
r^{-1}\int_{r_{b}}^{r_{e}}B_{i}(r')B_{j}(r'){\rm d}r', & r>r_{e},\\
r^{-1}\int_{r_{b}}^{r}B_{i}(r')B_{j}(r'){\rm d}r'+\int_{r}^{r_{e}}r'^{-1}B_{i}(r')B_{j}(r'){\rm d}r', & r_{b}<r<r_{e},\\
\int_{r_{b}}^{r_{e}}r'^{-1}B_{i}(r')B_{j}(r'){\rm d}r', & r<r_{b}.
\end{cases}\label{eq:Vcoul}
\end{equation}
\end{widetext} Like the two-electron integrals discussed above, the
in-element potential $r_{b}<r<r_{e}$ has to be evaluated by slices
at every radial quadrature point $(r_{0},r_{1},\dots,r_{n-1})$ \begin{widetext}
\begin{align}
\int_{r_{b}}^{r_{k-1}}B_{i}(r')B_{j}(r'){\rm d}r' & =\int_{r_{b}}^{r_{0}}B_{i}(r')B_{j}(r'){\rm d}r'+\sum_{l=1}^{k-1}\int_{r_{l-1}}^{r_{l}}B_{i}(r')B_{j}(r'){\rm d}r'\label{eq:Vcoul-in1}\\
\int_{r}^{r_{e}}r'^{-1}B_{i}(r')B_{j}(r'){\rm d}r' & =\int_{r_{n-1}}^{r_{e}}r'^{-1}B_{i}(r')B_{j}(r'){\rm d}r'+\sum_{l=k}^{n-1}\int_{r_{l-1}}^{r_{l}}r'^{-1}B_{i}(r')B_{j}(r'){\rm d}r'\label{eq:Vcoul-in2}
\end{align}
\end{widetext}

\subsubsection{Exchange-correlation potential\label{subsec:Exchange-correlation-potential}}

The functional derivative satisfies

\begin{equation}
\delta E=E[n+\delta n]-E[n]=\int\frac{\delta E}{\delta n}\delta n{\rm d}^{3}r.\label{eq:dExc}
\end{equation}
and so
\begin{equation}
\delta E=\int\left(\frac{\delta E}{\delta n}\delta n+\frac{\delta E}{\delta\nabla n}\delta\nabla n\right){\rm d}^{3}r.\label{eq:dExc-2}
\end{equation}
Integrating by parts one gets
\begin{align}
\int\frac{\delta E}{\delta\nabla n}\delta\nabla n{\rm d}^{3}r & =\left[\frac{\delta E}{\delta\nabla n}\delta n{\rm d}^{3}r-\int\nabla\frac{\delta E}{\delta\nabla n}\delta n{\rm d}^{3}r\right]\label{eq:dExc-part}\\
 & =-\int\nabla\frac{\delta E}{\delta\nabla n}\delta n{\rm d}^{3}r
\end{align}
from which one can identify
\begin{equation}
v(\boldsymbol{r})=\frac{\delta E}{\delta n}-\nabla\frac{\delta E}{\delta\nabla n}.\label{eq:vxc-GGA}
\end{equation}
Expressing the functional in terms of
\begin{equation}
\gamma^{\sigma\sigma'}=\nabla n^{\sigma}\cdot\nabla n^{\sigma'}\label{eq:redgrad}
\end{equation}
one has
\begin{equation}
\frac{\delta}{\delta\nabla n}=\frac{\delta\gamma}{\delta\nabla n}\frac{\delta}{\delta\gamma}=2\nabla n\label{eq:dgn}
\end{equation}
and so
\begin{equation}
v(\boldsymbol{r})=\frac{\delta E}{\delta n}-2\nabla\cdot\left(\frac{\delta E}{\delta\gamma}\nabla n\right)\label{eq:vxc-dgn}
\end{equation}
or for an open shell system
\begin{equation}
v^{\sigma}(\boldsymbol{r})=\frac{\delta E}{\delta n^{\sigma}}-\nabla\cdot\left(2\frac{\delta E}{\delta\gamma^{\sigma\sigma}}\nabla n^{\sigma}+\frac{\delta E}{\delta\gamma^{\sigma\sigma'}}\nabla n^{\sigma'}\right)\label{eq:vxc-dgn-os}
\end{equation}
where $\sigma\neq\sigma'$.

To guarantee accuracy, the gradient terms have to be evaluated analytically.
Fortunately, there's only radial dependence, so the gradient 
\begin{equation}
\nabla f=\left(\frac{\partial f}{\partial r},0,0\right)\label{eq:gn-rad}
\end{equation}
can be replaced by a radial derivative, and the divergence with
\begin{align}
\nabla\cdot\boldsymbol{A} & =\left(\frac{1}{r^{2}}\frac{\partial}{\partial r}\left(r^{2}A_{r}\right),0,0\right)=\left(\frac{\partial}{\partial r}A_{r}+\frac{2A_{r}}{r},0,0\right)\label{eq:divA}
\end{align}
Now,
\begin{align}
\partial_{r}\left(\frac{\delta E}{\delta\gamma^{\sigma\sigma'}}\partial_{r}n^{\sigma'}\right)= & \left(\partial_{r}\frac{\delta E}{\delta\gamma^{\sigma\sigma'}}\right)\cdot\partial_{r}n^{\sigma'}+\frac{\delta E}{\delta\gamma^{\sigma\sigma'}}\partial_{r}^{2}n^{\sigma'}\label{eq:vxc-1}
\end{align}
where

\begin{align}
\partial_{r}\left(\frac{\delta E}{\delta\gamma^{\sigma\sigma'}}\right) & =\frac{\partial n^{\tau}}{\partial r}\left(\frac{\partial}{\partial n^{\tau}}\frac{\delta E}{\delta\gamma^{\sigma\sigma'}}\right)+\frac{\partial\gamma^{\tau\tau'}}{\partial r}\left(\frac{\partial}{\partial\gamma^{\tau\tau'}}\frac{\delta E}{\delta\gamma^{\sigma\sigma'}}\right)\nonumber \\
 & =g^{\tau}\frac{\delta^{2}E}{\delta n^{\tau}\delta\gamma^{\sigma\sigma'}}+\left(l^{\tau}g^{\tau'}+g^{\tau}l^{\tau'}\right)\frac{\delta^{2}E}{\delta\gamma^{\tau\tau'}\delta\gamma^{\sigma\sigma'}}\label{eq:vxc-2}
\end{align}
where we have defined $g^{\tau}=\partial_{r}n^{\tau}$ and $l^{\tau}=\partial_{r}^{2}n^{\tau}$,
 and the extra $2A_{r}/r$ term from the divergence, \eqref{divA},
yielding
\begin{equation}
\frac{2}{r}\left(2\frac{\delta E}{\delta\gamma^{\sigma\sigma}}g^{\sigma}+\frac{\delta E}{\delta\gamma^{\sigma\sigma'}}g^{\sigma'}\right).\label{eq:vxc-3}
\end{equation}
Thus, altogether, the radial exchange(-correlation) potential is given
by\begin{widetext}
\begin{align}
v_{xc}^{\sigma}(r) & =\frac{\delta E}{\delta n^{\sigma}}-\nabla\cdot\left(2\frac{\delta E}{\delta\gamma^{\sigma\sigma}}\nabla n^{\sigma}+\frac{\delta E}{\delta\gamma^{\sigma\sigma'}}\nabla n^{\sigma'}\right)\label{eq:vxc-final}\\
 & =v_{\text{LDA}}^{\sigma}(r)-\frac{2}{r}\left[2\frac{\delta E}{\delta\gamma^{\sigma\sigma}}g^{\sigma}+\frac{\delta E}{\delta\gamma^{\sigma\sigma'}}g^{\sigma'}\right]\nonumber \\
 & -2\left[g^{\tau}\frac{\delta^{2}E}{\delta n^{\tau}\delta\gamma^{\sigma\sigma}}g^{\sigma}+\left(l^{\tau}g^{\tau'}+g^{\tau}l^{\tau'}\right)\frac{\delta^{2}E}{\delta\gamma^{\tau\tau'}\delta\gamma^{\sigma\sigma}}g^{\sigma}+\frac{\delta E}{\delta\gamma^{\sigma\sigma}}l^{\sigma}\right]\nonumber \\
 & -\left[g^{\tau}\frac{\delta^{2}E}{\delta n^{\tau}\delta\gamma^{\sigma\sigma'}}g^{\sigma'}+\left(l^{\tau}g^{\tau'}+g^{\tau}l^{\tau'}\right)\frac{\delta^{2}E}{\delta\gamma^{\tau\tau'}\delta\gamma^{\sigma\sigma'}}g^{\sigma'}+\frac{\delta E}{\delta\gamma^{\sigma\sigma'}}l^{\sigma'}\right]\label{eq:vxc-prod}
\end{align}
\end{widetext} where the various derivatives of the exchange-correlation
functional are available in \libxc{} \citep{Lehtola2018}.

\section{Results\label{sec:Results}}

To demonstrate the new routines, we reproduce literature values for
the ground states of the neutral and cationic atoms $1\leq Z\leq86$
with the VWN functional, as well as a Perdew--Burke--Ernzerhof (PBE)
\citep{Perdew1996,Perdew1997} functional based on VWN correlation
\citep{Kraisler2010}, accurate values for which were obtained in
\citeref{Kraisler2010} with 10~000 radial grid points. We found
that by using a radial basis consisting of ten 15-node elements and
a practical infinity $r_{\infty}=40a_{0}$, the energy has converged
beyond microhartree accuracy, even though this basis contains just
139 radial basis functions -- almost two orders of magnitude fewer
degrees of freedom than used in \citeref{Kraisler2010}. These results,
with differences to the reference data from \citeref{Kraisler2010}
are shown in \tabref{Total-energies-neut} for neutral atoms and \tabref{Total-energies-cat}
for their cations. 

The agreement to most part is excellent: large positive differences
that indicate the value of \citeref{Kraisler2010} is lower are seen
for the species for which the calculations in \citeref{Kraisler2010}
transferred fractional charge across shells. Otherwise, the differences
become noticeable for heavier atoms, nearing $-10\,\mu E_{h}$ when
$Z\to86$, indicating that the data in \citeref{Kraisler2010} is
not fully converged to the basis set limit. In our VWN calculation
on the \ce{La+} cation, it was discovered that the energy for the
$4f_{0}^{0.945}5d_{0}^{0.575}6s_{0}^{0.480}$ state reported in \citeref{Kraisler2010}
was incorrect; the correct energy is $-8217.456974$ \bibnote{Eli Kraisler, private communication, 2019.}.

\begin{table*}
\begin{tabular}{l|r@{\extracolsep{0pt}.}lrr@{\extracolsep{0pt}.}lr||l|r@{\extracolsep{0pt}.}lrr@{\extracolsep{0pt}.}lr}
$Z$ & \multicolumn{2}{c}{$E(\text{VWN})$} & $\Delta E(\text{VWN})$ & \multicolumn{2}{c}{$E(\text{PBE})$} & $\Delta E(\text{PBE})$ & $Z$ & \multicolumn{2}{c}{$E(\text{VWN})$} & $\Delta E(\text{VWN})$ & \multicolumn{2}{c}{$E(\text{PBE})$} & $\Delta E(\text{PBE})$\tabularnewline
\hline 
\hline 
 1 & -0&478671 &  0 & -0&499963 & -0 & 44 & -4439&044607 &  1 & -4443&255631 & -6\tabularnewline
 2 & -2&834836 &  0 & -2&893287 & -1 & 45 & -4683&360538 &  1 & -4687&685035 & -6\tabularnewline
 3 & -7&343957 &  0 & -7&462726 & -0 & 46 & -4935&368406 &  0 & -4939&811859 & -5\tabularnewline
 4 & -14&447209 & -0 & -14&630525 & -0 & 47 & -5195&037351 &  1 & -5199&600560 & -6\tabularnewline
 5 & -24&353614 &  0 & -24&606283 & -0 & 48 & -5462&390982 &  1 & -5467&070748 & -5\tabularnewline
 6 & -37&470031 &  0 & -37&795116 & -0 & 49 & -5737&313809 & -0 & -5742&114862 & -6\tabularnewline
 7 & -54&136799 &  0 & -54&537743 & -0 & 50 & -6019&972345 &  0 & -6024&895821 & -6\tabularnewline
 8 & -74&527410 &  0 & -75&003219 & -1 & 51 & -6310&419326 &  1 & -6315&467009 & -6\tabularnewline
 9 & -99&114192 &  0 & -99&668342 & -1 & 52 & -6608&650476 &  0 & -6613&811930 & -6\tabularnewline
10 & -128&233481 & -0 & -128&869661 & -1 & 53 & -6914&777857 &  1 & -6920&056261 & -6\tabularnewline
11 & -161&447625 & -0 & -162&176267 & -1 & 54 & -7228&856106 &  1 & -7234&254478 & -7\tabularnewline
12 & -199&139406 & -0 & -199&958820 & -1 & 55 & -7550&561866 &  0 & -7556&083016 & -7\tabularnewline
13 & -241&321156 & -0 & -242&236076 & -1 & 56 & -7880&111578 & -0 & -7885&752916 & -7\tabularnewline
14 & -288&222945 &  0 & -289&236535 & -1 & 57 & -8217&648931 &  91 & -8223&406822 &  220\tabularnewline
15 & -340&005794 & -0 & -341&120757 & -2 & 58 & -8563&489711 &  880 & -8569&364450 &  663\tabularnewline
16 & -396&743948 &  0 & -397&951200 & -1 & 59 & -8917&715777 &  6966 & -8923&707847 &  6539\tabularnewline
17 & -458&671463 & -0 & -459&976078 & -2 & 60 & -9280&405670 &  16518 & -9286&515623 &  16052\tabularnewline
18 & -525&946195 &  0 & -527&352025 & -2 & 61 & -9651&650420 &  13676 & -9657&878495 &  13601\tabularnewline
19 & -598&206032 & -0 & -599&716752 & -2 & 62 & -10031&516930 &  6437 & -10037&864135 &  6383\tabularnewline
20 & -675&742283 &  0 & -677&355243 & -2 & 63 & -10420&023146 &  1 & -10426&490411 & -8\tabularnewline
21 & -758&685248 &  0 & -760&397795 & -3 & 64 & -10817&148260 &  858 & -10823&727509 &  285\tabularnewline
22 & -847&314902 & -0 & -849&129808 & -3 & 65 & -11223&108037 &  4860 & -11229&800083 &  3513\tabularnewline
23 & -941&786662 & -0 & -943&704413 & -2 & 66 & -11637&977781 &  11030 & -11644&783485 &  9215\tabularnewline
24 & -1042&218348 & -0 & -1044&239902 & -3 & 67 & -12061&832318 &  18521 & -12068&752563 &  16445\tabularnewline
25 & -1148&644093 &  0 & -1150&765417 & -3 & 68 & -12494&746152 &  26769 & -12501&781831 &  24564\tabularnewline
26 & -1261&223291 &  6017 & -1263&441835 &  4291 & 69 & -12936&809752 &  19185 & -12943&957962 &  20751\tabularnewline
27 & -1380&193787 &  716 & -1382&508399 &  831 & 70 & -13388&048594 &  1 & -13395&317842 & -9\tabularnewline
28 & -1505&672905 &  1 & -1508&087914 & -4 & 71 & -13848&234767 &  1 & -13855&623680 & -9\tabularnewline
29 & -1637&793358 &  0 & -1640&310279 & -4 & 72 & -14317&517965 &  2 & -14325&032671 & -9\tabularnewline
30 & -1776&573850 &  0 & -1779&194575 & -4 & 73 & -14795&971453 &  1 & -14803&612704 & -9\tabularnewline
31 & -1921&851924 & -0 & -1924&582672 & -3 & 74 & -15283&610347 &  2 & -15291&380462 & -9\tabularnewline
32 & -2073&829860 &  0 & -2076&672928 & -4 & 75 & -15780&381133 &  2 & -15788&268506 & -8\tabularnewline
33 & -2232&587154 &  0 & -2235&545023 & -4 & 76 & -16286&434007 &  2 & -16294&440422 & -9\tabularnewline
34 & -2398&134930 &  0 & -2401&196896 & -4 & 77 & -16801&850893 &  2 & -16809&976281 & -8\tabularnewline
35 & -2570&626651 & -0 & -2573&796934 & -4 & 78 & -17326&660985 &  3 & -17334&912620 & -10\tabularnewline
36 & -2750&147940 &  1 & -2753&430126 & -4 & 79 & -17860&796573 &  2 & -17869&175326 & -10\tabularnewline
37 & -2936&342160 &  0 & -2939&739646 & -5 & 80 & -18404&274220 &  1 & -18412&777007 & -10\tabularnewline
38 & -3129&453161 &  1 & -3132&963153 & -5 & 81 & -18956&962102 &  1 & -18965&593468 & -10\tabularnewline
39 & -3329&525142 &  0 & -3333&148098 & -5 & 82 & -19519&010773 &  2 & -19527&771776 & -10\tabularnewline
40 & -3536&771074 & -0 & -3540&515940 & -5 & 83 & -20090&453943 &  1 & -20099&346370 & -9\tabularnewline
41 & -3751&295618 &  0 & -3755&160742 & -6 & 84 & -20671&273855 &  2 & -20680&287630 & -9\tabularnewline
42 & -3973&162595 & -0 & -3977&149787 & -5 & 85 & -21261&559507 &  2 & -21270&697436 & -10\tabularnewline
43 & -4202&348934 &  1 & -4206&446961 & -5 & 86 & -21861&346869 &  3 & -21870&611766 & -9\tabularnewline
\end{tabular}

\caption{Total energies of neutral atoms for calculations with the VWN functional
and a PBE functional based on VWN correlation. The differences $\Delta E=E(\text{present work)}-E(\text{Kraisler 2010})$
are calculated relative to the fully numerical values from \citeref{Kraisler2010}
and are reported in microhartree. \label{tab:Total-energies-neut}}
\end{table*}

\begin{table*}
\begin{tabular}{l|r@{\extracolsep{0pt}.}lrr@{\extracolsep{0pt}.}lr||l|r@{\extracolsep{0pt}.}lrr@{\extracolsep{0pt}.}lr}
$Z$ & \multicolumn{2}{c}{$E(\text{VWN})$} & $\Delta E(\text{VWN})$ & \multicolumn{2}{c}{$E(\text{PBE})$} & $\Delta E(\text{PBE})$ & $Z$ & \multicolumn{2}{c}{$E(\text{VWN})$} & $\Delta E(\text{VWN})$ & \multicolumn{2}{c}{$E(\text{PBE})$} & $\Delta E(\text{PBE})$\tabularnewline
\hline 
\hline 
 1 &  0&000000 &  0 &  0&000000 &  0 & 44 & -4438&767375 &  1 & -4442&987586 & -6\tabularnewline
 2 & -1&941703 &  0 & -1&993741 & -1 & 45 & -4683&055688 & -0 & -4687&391656 & -6\tabularnewline
 3 & -7&142818 & -0 & -7&257274 & -0 & 46 & -4935&023835 & -0 & -4939&477807 & -6\tabularnewline
 4 & -14&115512 & -0 & -14&299957 & -0 & 47 & -5194&755725 &  0 & -5199&329663 & -6\tabularnewline
 5 & -24&038275 & -0 & -24&294185 & -0 & 48 & -5462&065582 &  1 & -5466&758188 & -6\tabularnewline
 6 & -37&037413 &  0 & -37&365761 & -1 & 49 & -5737&101461 &  1 & -5741&909394 & -6\tabularnewline
 7 & -53&585407 & -0 & -53&989486 & -1 & 50 & -6019&697098 &  0 & -6024&624904 & -6\tabularnewline
 8 & -74&016721 & -0 & -74&500466 & -1 & 51 & -6310&085269 &  0 & -6315&135445 & -6\tabularnewline
 9 & -98&450427 &  0 & -99&012379 & -1 & 52 & -6608&317272 &  0 & -6613&492318 & -6\tabularnewline
10 & -127&418114 & -0 & -128&061506 & -1 & 53 & -6914&378893 &  1 & -6919&666895 & -6\tabularnewline
11 & -161&250340 &  0 & -161&979238 & -2 & 54 & -7228&394173 &  0 & -7233&799075 & -7\tabularnewline
12 & -198&855669 &  0 & -199&679196 & -1 & 55 & -7550&416737 &  0 & -7555&942187 & -7\tabularnewline
13 & -241&100595 &  0 & -242&016828 & -2 & 56 & -7879&920522 &  984 & -7885&569221 &  813\tabularnewline
14 & -287&918773 & -0 & -288&932217 & -1 & 57$^{a}$ & -8217&455429 &  1545 & -8223&221428 &  2463\tabularnewline
15 & -339&618451 & -0 & -340&732787 & -2 & 58 & -8563&294377 &  7617 & -8569&175665 &  6993\tabularnewline
16 & -396&356540 &  0 & -397&574834 & -1 & 59 & -8917&518819 &  20242 & -8923&517325 &  19853\tabularnewline
17 & -458&184562 &  0 & -459&496045 & -2 & 60 & -9280&230184 &  12512 & -9286&344462 &  14304\tabularnewline
18 & -525&360439 & -0 & -526&770845 & -2 & 61 & -9651&482296 &  5894 & -9657&715083 &  7570\tabularnewline
19 & -598&039506 & -0 & -599&553203 & -2 & 62 & -10031&348348 &  2185 & -10037&700262 &  3563\tabularnewline
20 & -675&514035 & -0 & -677&132344 & -3 & 63 & -10419&820399 &  1 & -10426&294380 & -8\tabularnewline
21 & -758&442642 &  1812 & -760&161910 &  1575 & 64 & -10816&942313 &  782 & -10823&528176 & -8\tabularnewline
22 & -847&065015 &  2847 & -848&886280 &  2949 & 65 & -11222&899249 &  11493 & -11229&597850 &  8739\tabularnewline
23 & -941&523838 &  0 & -943&448454 & -2 & 66 & -11637&769436 &  20564 & -11644&578605 &  20710\tabularnewline
24 & -1041&944126 &  0 & -1043&972993 & -3 & 67 & -12061&641951 &  12969 & -12068&562285 &  16381\tabularnewline
25 & -1148&368924 & -0 & -1150&502293 & -3 & 68 & -12494&571851 &  7644 & -12501&608100 &  10480\tabularnewline
26 & -1260&927746 & -0 & -1263&157817 & -3 & 69 & -12936&633640 &  4018 & -12943&786682 &  5267\tabularnewline
27 & -1379&896444 &  0 & -1382&218969 & -3 & 70 & -13387&827982 &  1 & -13395&103732 & -10\tabularnewline
28 & -1505&370040 &  0 & -1507&793428 & -4 & 71 & -13847&999554 &  2 & -13855&396289 & -9\tabularnewline
29 & -1637&485140 &  0 & -1640&010907 & -4 & 72 & -14317&267886 &  1533 & -14324&788056 &  2091\tabularnewline
30 & -1776&217890 &  0 & -1778&850041 & -3 & 73 & -14795&705335 &  1 & -14803&354716 & -8\tabularnewline
31 & -1921&629140 & -0 & -1924&365546 & -4 & 74 & -15283&334783 &  2 & -15291&113757 & -9\tabularnewline
32 & -2073&533337 &  0 & -2076&379628 & -4 & 75 & -15780&100605 &  2 & -15788&005290 & -8\tabularnewline
33 & -2232&220332 &  0 & -2235&179888 & -4 & 76 & -16286&150952 &  2 & -16294&166211 & -8\tabularnewline
34 & -2397&770127 & -0 & -2400&845945 & -4 & 77 & -16801&535551 &  3 & -16809&673022 & -8\tabularnewline
35 & -2570&180737 & -0 & -2573&360358 & -4 & 78 & -17326&305178 &  1 & -17334&567568 & -10\tabularnewline
36 & -2749&623528 &  1 & -2752&911995 & -4 & 79 & -17860&511437 &  0 & -17868&901185 & -10\tabularnewline
37 & -2936&183045 &  0 & -2939&584557 & -5 & 80 & -18403&949940 &  1 & -18412&465847 & -10\tabularnewline
38 & -3129&240801 & -0 & -3132&756842 & -5 & 81 & -18956&753577 &  1 & -18965&392384 & -10\tabularnewline
39 & -3329&295616 & -0 & -3332&926478 & -5 & 82 & -19518&743995 &  2 & -19527&509994 & -10\tabularnewline
40 & -3536&524561 &  871 & -3540&277393 &  131 & 83 & -20090&133387 &  2 & -20099&028863 & -9\tabularnewline
41 & -3751&036938 &  0 & -3754&910732 & -5 & 84 & -20670&954321 &  2 & -20679&981706 & -9\tabularnewline
42 & -3972&894262 & -0 & -3976&890719 & -5 & 85 & -21261&180969 &  2 & -21270&328813 & -8\tabularnewline
43 & -4202&074923 &  0 & -4206&181883 & -5 & 86 & -21860&912366 &  3 & -21870&184141 & -9\tabularnewline
\end{tabular}

\caption{Total energies of atomic cations for calculations with the VWN functional
and a PBE functional based on VWN correlation. The differences $\Delta E=E(\text{present work)}-E(\text{Kraisler 2010})$
are calculated relative to the fully numerical values from \citeref{Kraisler2010}
and are reported in microhartree. $^{a}$An incorrect value was reported
in \citeref{Kraisler2010} for \ce{La+}; see main text. \label{tab:Total-energies-cat}}
\end{table*}

Next, we demonstrate that the erfc range-separation scheme works by
reproducing literature values for the total energies of the spherically
symmetric atoms on the first two rows using the long-range corrected
BLYP functional \citep{Iikura2001,Lee1988,Anderson2017}. In \citeref{Lehtola2019a}
we investigated the accuracy of the \apcinfty{} Gaussian basis set
that was used in \citeref{Anderson2017}. The study was restricted
to $^{n}S$ states to avoid symmetry breaking effects, which were
still observed for \ce{H-}, He, \ce{Li-}, and \ce{Na-}, as was
discussed in the Introduction. Reproducing symmetry preserving data
with \Erkale{} \citep{Lehtola2012}, we found that the truncation
error of the \apcinfty{} basis set is less than 1 $\mu E_{h}$ for
light atoms and tens of $\mu E_{h}$ for heavier atoms in Hartree--Fock
and BHHLYP \citep{Becke1993a} calculations.

Because the screening is evaluated analytically in Gaussian-basis
calculations \citep{Heyd2003} and the accuracy of the \apcinfty{}
basis set has been established \citep{Lehtola2019a}, the values reported
in \citeref{Anderson2017} offer an ideal reference for the present
work. The comparison of results obtained in the present work with
\eqrangeref{phin}{Dnk} and a numerical basis set with five 15-node
radial elements and a practical infinity $r_{\infty}=40a_{0}$ is
shown in \tabref{Comparison-of-the}, demonstrating excellent agreement
between the fully numerical and Gaussian basis calculations. The values
are in full agreement after rounding to the same accuracy for the
light atoms, while the fully numerical values are slightly below the
Gaussian-basis values for the heavier atoms, as expected based on
the basis set truncation errors observed in \citeref{Lehtola2019a}.

\begin{table}
\begin{tabular}{l|r@{\extracolsep{0pt}.}lr@{\extracolsep{0pt}.}l}
atom & \multicolumn{2}{c}{finite element} & \multicolumn{2}{c}{Gaussian basis}\tabularnewline
\hline 
\hline 
\ce{H-} &    -0&519949 & -0&51995\tabularnewline
He &    -2&866811 & -2&86681\tabularnewline
\ce{Li-} &    -7&435511 & -7&43551\tabularnewline
Be &   -14&584723 & -14&58472\tabularnewline
N &   -54&482223 & -54&48222\tabularnewline
\ce{F-} & -99&766050 & -99&76604\tabularnewline
Ne &  -128&816627 & -128&81661\tabularnewline
\ce{Na-} &  -162&136564 & -162&13655\tabularnewline
Mg &  -199&907036 & -199&90702\tabularnewline
P & -341&069932 & -341&06992\tabularnewline
\ce{Cl-} & -460&080588 & -460&08057\tabularnewline
Ar &  -527&321257 & -527&32124\tabularnewline
\end{tabular}

\caption{Comparison of the total energies of spherically symmetric atoms with
the LC-BLYP functional with the range separation constant $\omega=0.3$
reproduced with finite element calculations (present work) and a Gaussian
basis set calculation \citep{Anderson2017}. \label{tab:Comparison-of-the}}
\end{table}

Finally, the spin-restricted ground states for all atoms in the periodic
table at HF and HFS levels of theory are shown in \tabref{HF-gs, HFS-gs},
respectively; these calculations also used ten 15-node radial elements.
The data reveal that in some cases a lower-lying configuration has
been seen in the brute force search, but that it failed to converge.
In the HF calculations, the $6s^{2}4f^{2}5d^{1}$ state of Pr converges
without problems; however, the $6s^{2}4f^{3}$ configuration has a
lower energy but its wave function failed to converge. Similar issues
are observed in the HFS calculations for Cf, Es, and Fm, where the
$5f^{n-1}7s^{1}$ state converges without problems, but a lower energy
is observed for a $5f^{n}$ configuration the wave function of which
fails to converge.

The HF results can be compared to the high-accuracy data for multiconfigurational
HF of Saito \citep{Saito2009}. Because the present calculations are
spin-restricted with fractional occupations, the energies are higher
than those reported in \citeref{Saito2009}. However, the agreement
for the noble gases is perfect, underlining the high accuracy of the
computational approach used in the present work, which was outlined
in \citeref{Lehtola2019a}, even though only 139 radial basis functions
were employed.

\begin{table*}
\small
\input{HF.tex}

\caption{Non-relativistic spin-restricted spherical HF configurations for all
elements in the periodic table. Entries in italic indicate a lower-lying
configuration was identified but it failed to converge.\label{tab:HF-gs}}

\end{table*}

\begin{table*}
\small
\input{LDAX.tex}

\caption{Non-relativistic spin-restricted spherical HFS configurations for
all elements in the periodic table. Entries in italic indicate a lower-lying
configuration was identified but it failed to converge.\label{tab:HFS-gs}}
\end{table*}

\section{Summary and discussion\label{sec:Summary-and-discussion}}

We have described new efficient implementations of range-separated
functionals as well as fractional occupations for atomic electronic
structure calculations with \HelFEM{}, and demonstrated that beyond
microhartree accucacy can be achieved with just 139 numerical radial
basis functions. We have tested the program by reproducing local density
approximation (LDA) and generalized gradient approximation (GGA) total
energies for $1\leq Z\leq86$ at the basis set limit, and shown that
the literature values deviate from the complete basis set limit by
up to $10\ \mu E_{h}$. The approaches developed in the present work
could be straightforwardly extended to fractional occupations per
shell in future work, requiring the addition of a logic to formulate
the fractional occupation numbers.

The capabilities added to \HelFEM{} in the present work allow for
self-consistent benchmarking of density functionals at the basis set
limit, which is useful for development and implementation purposes.
For instance, Clementi--Roetti wave functions \citep{Clementi1974}
are often used for non-self-consistent benchmarks of density functionals,
but the availability of a program for self-consistent calculations
is certain to help future developments as numerical instabilities
in the functional may not be detected in non-self-consistent calculations.

Furthermore, we have reported the non-relativistic spin-restricted
ground state configurations of all atoms in the periodic table at
HF and HFS levels of theory. Such knowledge is useful for implementations
of the superposition of atomic densities guess \citep{Almlof1982,VanLenthe2006},
which is often implemented based on spin-restricted fractionally occupied
calculations. The present approach is also useful for implementations
of the SAP guess \citep{Lehtola2019}. For instance, the implementation
of SAP now available in the development version of the \PsiFour{}
program \citep{Parrish2017} is based on HFS potentials tabulated
during the present work. Instead of the 4000 point tabulation used
in \citeref{Lehtola2019} with unknown error, the ten-element calculations
of the present work yield 751-point tabulations that reproduce the
sub-microhartree-level accuracy of the original calculation.

The atomic orbitals obtained from the present approach may also be
useful for initializing fully numerical molecular electronic structure
calculations via either a superposition of atomic densities, or in
combination to the extended Hückel rule developed in \citeref{Lehtola2019}.

\section*{Funding information}

This work has been supported by the Academy of Finland through project
number 311149.

\section*{Acknowledgments}

I thank Dirk Andrae, Volker Blum, Eli Kraisler, Jacek Kobus, Leeor
Kronik, Micael Oliveira, Dage Sundholm, Edward Valeev, and Lucas Visscher
for discussions. Computational resources provided by CSC -- It Center
for Science Ltd (Espoo, Finland) and the Finnish Grid and Cloud Infrastructure
(persistent identifier urn:nbn:fi:research-infras-2016072533) are
gratefully acknowledged.

\bibliography{citations}

\end{document}

%% file: HF.tex
\begin{tabular}{llr|llr|llr}
H & $ 1s^{1} $ & -0.357710 & Nb & $ \textrm{[Kr]} 5s^{2} 4d^{3} $ & -3753.033166 & Tl & $ \textrm{[Xe]} 4f^{14} 5d^{10} 6s^{2} 6p^{1} $ & -18961.740923\tabularnewline
He & $ 1s^{2} $ & -2.861680 & Mo & $ \textrm{[Kr]} 4d^{6} $ & -3974.815043 & Pb & $ \textrm{[Xe]} 4f^{14} 5d^{10} 6s^{2} 6p^{2} $ & -19523.831389\tabularnewline
Li & $ \textrm{[He]} 2s^{1} $ & -7.378133 & Tc & $ \textrm{[Kr]} 4d^{7} $ & -4204.141902 & Bi & $ \textrm{[Xe]} 4f^{14} 5d^{10} 6s^{2} 6p^{3} $ & -20095.328128\tabularnewline
Be & $ \textrm{[He]} 2s^{2} $ & -14.573023 & Ru & $ \textrm{[Kr]} 4d^{8} $ & -4441.038680 & Po & $ \textrm{[Xe]} 4f^{14} 5d^{10} 6s^{2} 6p^{4} $ & -20676.283998\tabularnewline
B & $ \textrm{[He]} 2s^{2} 2p^{1} $ & -24.384693 & Rh & $ \textrm{[Kr]} 4d^{9} $ & -4685.600291 & At & $ \textrm{[Xe]} 4f^{14} 5d^{10} 6s^{2} 6p^{5} $ & -21266.749081\tabularnewline
C & $ \textrm{[He]} 2s^{2} 2p^{2} $ & -37.344157 & Pd & $ \textrm{[Kr]} 4d^{10} $ & -4937.921024 & Rn & $ \textrm{[Xe]} 4f^{14} 5d^{10} 6s^{2} 6p^{6} $ & -21866.772241\tabularnewline
N & $ \textrm{[He]} 2s^{2} 2p^{3} $ & -53.852155 & Ag & $ \textrm{[Kr]} 4d^{10} 5s^{1} $ & -5197.639939 & Fr & $ \textrm{[Rn]} 7s^{1} $ & -22475.826522\tabularnewline
O & $ \textrm{[He]} 2s^{2} 2p^{4} $ & -74.297532 & Cd & $ \textrm{[Kr]} 4d^{10} 5s^{2} $ & -5465.133143 & Ra & $ \textrm{[Rn]} 7s^{2} $ & -23094.303666\tabularnewline
F & $ \textrm{[He]} 2s^{2} 2p^{5} $ & -99.067145 & In & $ \textrm{[Kr]} 4d^{10} 5s^{2} 5p^{1} $ & -5740.082317 & Ac & $ \textrm{[Rn]} 7s^{2} 6d^{1} $ & -23722.073196\tabularnewline
Ne & $ \textrm{[He]} 2s^{2} 2p^{6} $ & -128.547098 & Sn & $ \textrm{[Kr]} 4d^{10} 5s^{2} 5p^{2} $ & -6022.746221 & Th & $ \textrm{[Rn]} 7s^{2} 6d^{2} $ & -24359.362900\tabularnewline
Na & $ \textrm{[Ne]} 3s^{1} $ & -161.808533 & Sb & $ \textrm{[Kr]} 4d^{10} 5s^{2} 5p^{3} $ & -6313.211503 & Pa & $ \textrm{[Rn]} 7s^{2} 5f^{3} $ & -25006.406325\tabularnewline
Mg & $ \textrm{[Ne]} 3s^{2} $ & -199.614636 & Te & $ \textrm{[Kr]} 4d^{10} 5s^{2} 5p^{4} $ & -6611.551696 & U & $ \textrm{[Rn]} 7s^{2} 5f^{4} $ & -25663.398242\tabularnewline
Al & $ \textrm{[Ne]} 3s^{2} 3p^{1} $ & -241.782323 & I & $ \textrm{[Kr]} 4d^{10} 5s^{2} 5p^{5} $ & -6917.837495 & Np & $ \textrm{[Rn]} 7s^{2} 5f^{5} $ & -26330.321976\tabularnewline
Si & $ \textrm{[Ne]} 3s^{2} 3p^{2} $ & -288.637472 & Xe & $ \textrm{[Kr]} 4d^{10} 5s^{2} 5p^{6} $ & -7232.138364 & Pu & $ \textrm{[Rn]} 5f^{7} 7s^{1} $ & -27007.271797\tabularnewline
P & $ \textrm{[Ne]} 3s^{2} 3p^{3} $ & -340.381142 & Cs & $ \textrm{[Xe]} 6s^{1} $ & -7553.899845 & Am & $ \textrm{[Rn]} 5f^{9} $ & -27694.356363\tabularnewline
S & $ \textrm{[Ne]} 3s^{2} 3p^{4} $ & -397.202080 & Ba & $ \textrm{[Xe]} 6s^{2} $ & -7883.543827 & Cm & $ \textrm{[Rn]} 5f^{10} $ & -28391.573019\tabularnewline
Cl & $ \textrm{[Ne]} 3s^{2} 3p^{5} $ & -459.286063 & La & $ \textrm{[Xe]} 6s^{2} 5d^{1} $ & -8220.935691 & Bk & $ \textrm{[Rn]} 5f^{11} $ & -29098.977559\tabularnewline
Ar & $ \textrm{[Ne]} 3s^{2} 3p^{6} $ & -526.817513 & Ce & $ \textrm{[Xe]} 6s^{2} 5d^{2} $ & -8566.342481 & Cf & $ \textrm{[Rn]} 5f^{12} $ & -29816.624759\tabularnewline
K & $ \textrm{[Ar]} 4s^{1} $ & -599.124244 & Pr & $ \textit{[Xe]} 6s^{2} 4f^{2} 5d^{1} $ & {\em  -8920.094872} & Es & $ \textrm{[Rn]} 5f^{13} $ & -30544.570349\tabularnewline
Ca & $ \textrm{[Ar]} 4s^{2} $ & -676.758186 & Nd & $ \textrm{[Xe]} 6s^{2} 4f^{4} $ & -9282.434373 & Fm & $ \textrm{[Rn]} 5f^{14} $ & -31282.870930\tabularnewline
Sc & $ \textrm{[Ar]} 4s^{2} 4p^{1} $ & -759.556762 & Pm & $ \textrm{[Xe]} 6s^{2} 4f^{5} $ & -9653.359914 & Md & $ \textrm{[Rn]} 5f^{14} 7s^{1} $ & -32031.135295\tabularnewline
Ti & $ \textrm{[Ar]} 4s^{2} 3d^{2} $ & -847.933865 & Sm & $ \textrm{[Xe]} 6s^{2} 4f^{6} $ & -10032.949725 & No & $ \textrm{[Rn]} 5f^{14} 7s^{2} $ & -32789.512140\tabularnewline
V & $ \textrm{[Ar]} 4s^{2} 3d^{3} $ & -942.147322 & Eu & $ \textrm{[Xe]} 4f^{7} 6s^{2} $ & -10421.286649 & Lr & $ \textrm{[Rn]} 5f^{14} 7s^{2} 6d^{1} $ & -33557.812903\tabularnewline
Cr & $ \textrm{[Ar]} 4s^{2} 3d^{4} $ & -1042.342957 & Gd & $ \textrm{[Xe]} 6s^{1} 4f^{9} $ & -10818.487373 & Rf & $ \textrm{[Rn]} 5f^{14} 7s^{2} 6d^{2} $ & -34336.316816\tabularnewline
Mn & $ \textrm{[Ar]} 3d^{7} $ & -1148.803487 & Tb & $ \textrm{[Xe]} 4f^{11} $ & -11224.646666 & Db & $ \textrm{[Rn]} 5f^{14} 7s^{2} 6d^{3} $ & -35125.088022\tabularnewline
Fe & $ \textrm{[Ar]} 3d^{8} $ & -1261.579698 & Dy & $ \textrm{[Xe]} 4f^{12} $ & -11639.819030 & Sg & $ \textrm{[Rn]} 5f^{14} 6d^{6} $ & -35924.293864\tabularnewline
Co & $ \textrm{[Ar]} 3d^{9} $ & -1380.817569 & Ho & $ \textrm{[Xe]} 4f^{13} $ & -12064.074984 & Bh & $ \textrm{[Rn]} 5f^{14} 6d^{7} $ & -36733.871607\tabularnewline
Ni & $ \textrm{[Ar]} 3d^{10} $ & -1506.669759 & Er & $ \textrm{[Xe]} 4f^{14} $ & -12497.495312 & Hs & $ \textrm{[Rn]} 5f^{14} 6d^{8} $ & -37553.863992\tabularnewline
Cu & $ \textrm{[Ar]} 3d^{10} 4s^{1} $ & -1638.899667 & Tm & $ \textrm{[Xe]} 4f^{14} 6s^{1} $ & -12939.976389 & Mt & $ \textrm{[Rn]} 5f^{14} 6d^{9} $ & -38384.313942\tabularnewline
Zn & $ \textrm{[Ar]} 3d^{10} 4s^{2} $ & -1777.848116 & Yb & $ \textrm{[Xe]} 4f^{14} 6s^{2} $ & -13391.456193 & Ds & $ \textrm{[Rn]} 5f^{14} 6d^{10} $ & -39225.264332\tabularnewline
Ga & $ \textrm{[Ar]} 3d^{10} 4s^{2} 4p^{1} $ & -1923.166449 & Lu & $ \textrm{[Xe]} 4f^{14} 6s^{2} 6p^{1} $ & -13851.687533 & Rg & $ \textrm{[Rn]} 5f^{14} 6d^{10} 7s^{1} $ & -40076.301440\tabularnewline
Ge & $ \textrm{[Ar]} 3d^{10} 4s^{2} 4p^{2} $ & -2075.150884 & Hf & $ \textrm{[Xe]} 4f^{14} 6s^{2} 5d^{2} $ & -14320.929628 & Cn & $ \textrm{[Rn]} 5f^{14} 6d^{10} 7s^{2} $ & -40937.797856\tabularnewline
As & $ \textrm{[Ar]} 3d^{10} 4s^{2} 4p^{3} $ & -2233.924574 & Ta & $ \textrm{[Xe]} 4f^{14} 6s^{2} 5d^{3} $ & -14799.321729 & Nh & $ \textrm{[Rn]} 5f^{14} 6d^{10} 7s^{2} 7p^{1} $ & -41809.456590\tabularnewline
Se & $ \textrm{[Ar]} 3d^{10} 4s^{2} 4p^{4} $ & -2399.595885 & W & $ \textrm{[Xe]} 4f^{14} 5d^{6} $ & -15286.959470 & Fl & $ \textrm{[Rn]} 5f^{14} 6d^{10} 7s^{2} 7p^{2} $ & -42691.493680\tabularnewline
Br & $ \textrm{[Ar]} 3d^{10} 4s^{2} 4p^{5} $ & -2572.270918 & Re & $ \textrm{[Xe]} 4f^{14} 5d^{7} $ & -15783.943765 & Mc & $ \textrm{[Rn]} 5f^{14} 6d^{10} 7s^{2} 7p^{3} $ & -43583.961779\tabularnewline
Kr & $ \textrm{[Ar]} 3d^{10} 4s^{2} 4p^{6} $ & -2752.054977 & Os & $ \textrm{[Xe]} 4f^{14} 5d^{8} $ & -16290.259414 & Lv & $ \textrm{[Rn]} 5f^{14} 6d^{10} 7s^{2} 7p^{4} $ & -44486.902550\tabularnewline
Rb & $ \textrm{[Kr]} 5s^{1} $ & -2938.319660 & Ir & $ \textrm{[Xe]} 4f^{14} 5d^{9} $ & -16805.965623 & Ts & $ \textrm{[Rn]} 5f^{14} 6d^{10} 7s^{2} 7p^{5} $ & -45400.354767\tabularnewline
Sr & $ \textrm{[Kr]} 5s^{2} $ & -3131.545686 & Pt & $ \textrm{[Xe]} 4f^{14} 5d^{10} $ & -17331.121868 & Og & $ \textrm{[Rn]} 5f^{14} 6d^{10} 7s^{2} 7p^{6} $ & -46324.355815\tabularnewline
Y & $ \textrm{[Kr]} 5s^{2} 5p^{1} $ & -3331.559557 & Au & $ \textrm{[Xe]} 4f^{14} 5d^{10} 6s^{1} $ & -17865.342083 &  & & \tabularnewline
Zr & $ \textrm{[Kr]} 5s^{2} 4d^{2} $ & -3538.662298 & Hg & $ \textrm{[Xe]} 4f^{14} 5d^{10} 6s^{2} $ & -18408.991495 &  & & \tabularnewline
\end{tabular}

%% file: LDAX.tex
\begin{tabular}{llr|llr|llr}
H & $ 1s^{1} $ & -0.406534 & Nb & $ \textrm{[Kr]} 4d^{3} 5s^{2} $ & -3747.428127 & Tl & $ \textrm{[Xe]} 4f^{14} 5d^{10} 6s^{2} 6p^{1} $ & -18948.496862\tabularnewline
He & $ 1s^{2} $ & -2.723640 & Mo & $ \textrm{[Kr]} 5s^{1} 4d^{5} $ & -3969.125868 & Pb & $ \textrm{[Xe]} 4f^{14} 5d^{10} 6s^{2} 6p^{2} $ & -19510.422489\tabularnewline
Li & $ \textrm{[He]} 2s^{1} $ & -7.174881 & Tc & $ \textrm{[Kr]} 4d^{6} 5s^{1} $ & -4198.246878 & Bi & $ \textrm{[Xe]} 4f^{14} 5d^{10} 6s^{2} 6p^{3} $ & -20081.732046\tabularnewline
Be & $ \textrm{[He]} 2s^{2} $ & -14.223291 & Ru & $ \textrm{[Kr]} 4d^{8} $ & -4434.888516 & Po & $ \textrm{[Xe]} 4f^{14} 5d^{10} 6s^{2} 6p^{4} $ & -20662.460965\tabularnewline
B & $ \textrm{[He]} 2s^{2} 2p^{1} $ & -24.050406 & Rh & $ \textrm{[Kr]} 4d^{9} $ & -4679.115070 & At & $ \textrm{[Xe]} 4f^{14} 5d^{10} 6s^{2} 6p^{5} $ & -21252.645251\tabularnewline
C & $ \textrm{[He]} 2s^{2} 2p^{2} $ & -37.053605 & Pd & $ \textrm{[Kr]} 4d^{10} $ & -4931.010033 & Rn & $ \textrm{[Xe]} 4f^{14} 5d^{10} 6s^{2} 6p^{6} $ & -21852.321426\tabularnewline
N & $ \textrm{[He]} 2s^{2} 2p^{3} $ & -53.567903 & Ag & $ \textrm{[Kr]} 4d^{10} 5s^{1} $ & -5190.567420 & Fr & $ \textrm{[Rn]} 7s^{1} $ & -22461.201212\tabularnewline
O & $ \textrm{[He]} 2s^{2} 2p^{4} $ & -73.925425 & Cd & $ \textrm{[Kr]} 4d^{10} 5s^{2} $ & -5457.821825 & Ra & $ \textrm{[Rn]} 7s^{2} $ & -23079.470637\tabularnewline
F & $ \textrm{[He]} 2s^{2} 2p^{5} $ & -98.456607 & In & $ \textrm{[Kr]} 4d^{10} 5s^{2} 5p^{1} $ & -5732.640932 & Ac & $ \textrm{[Rn]} 7s^{2} 5f^{1} $ & -23707.189388\tabularnewline
Ne & $ \textrm{[He]} 2s^{2} 2p^{6} $ & -127.490741 & Sn & $ \textrm{[Kr]} 4d^{10} 5s^{2} 5p^{2} $ & -6015.182678 & Th & $ \textrm{[Rn]} 5f^{2} 7s^{2} $ & -24344.622650\tabularnewline
Na & $ \textrm{[Ne]} 3s^{1} $ & -160.628228 & Sb & $ \textrm{[Kr]} 4d^{10} 5s^{2} 5p^{3} $ & -6305.500906 & Pa & $ \textrm{[Rn]} 5f^{3} 7s^{2} $ & -24991.833379\tabularnewline
Mg & $ \textrm{[Ne]} 3s^{2} $ & -198.248792 & Te & $ \textrm{[Kr]} 4d^{10} 5s^{2} 5p^{4} $ & -6603.649656 & U & $ \textrm{[Rn]} 7s^{1} 5f^{5} $ & -25648.893676\tabularnewline
Al & $ \textrm{[Ne]} 3s^{2} 3p^{1} $ & -240.346857 & I & $ \textrm{[Kr]} 4d^{10} 5s^{2} 5p^{5} $ & -6909.683446 & Np & $ \textrm{[Rn]} 7s^{1} 5f^{6} $ & -26315.863733\tabularnewline
Si & $ \textrm{[Ne]} 3s^{2} 3p^{2} $ & -287.145287 & Xe & $ \textrm{[Kr]} 4d^{10} 5s^{2} 5p^{6} $ & -7223.657213 & Pu & $ \textrm{[Rn]} 7s^{1} 5f^{7} $ & -26992.780160\tabularnewline
P & $ \textrm{[Ne]} 3s^{2} 3p^{3} $ & -338.804261 & Cs & $ \textrm{[Xe]} 6s^{1} $ & -7545.272707 & Am & $ \textrm{[Rn]} 5f^{8} 7s^{1} $ & -27679.697021\tabularnewline
S & $ \textrm{[Ne]} 3s^{2} 3p^{4} $ & -395.481609 & Ba & $ \textrm{[Xe]} 6s^{2} $ & -7874.734118 & Cm & $ \textrm{[Rn]} 5f^{9} 7s^{1} $ & -28376.667807\tabularnewline
Cl & $ \textrm{[Ne]} 3s^{2} 3p^{5} $ & -457.333996 & La & $ \textrm{[Xe]} 6s^{2} 4f^{1} $ & -8212.148603 & Bk & $ \textrm{[Rn]} 5f^{10} 7s^{1} $ & -29083.745568\tabularnewline
Ar & $ \textrm{[Ne]} 3s^{2} 3p^{6} $ & -524.517426 & Ce & $ \textrm{[Xe]} 6s^{2} 4f^{2} $ & -8557.852692 & Cf & $ \textit{[Rn]} 5f^{11} 7s^{1} $ & {\em  -29800.983007}\tabularnewline
K & $ \textrm{[Ar]} 4s^{1} $ & -596.699051 & Pr & $ \textrm{[Xe]} 4f^{3} 6s^{2} $ & -8911.927706 & Es & $ \textit{[Rn]} 5f^{12} 7s^{1} $ & {\em  -30528.432552}\tabularnewline
Ca & $ \textrm{[Ar]} 4s^{2} $ & -674.160118 & Nd & $ \textrm{[Xe]} 4f^{4} 6s^{2} $ & -9274.451612 & Fm & $ \textit{[Rn]} 5f^{13} 7s^{1} $ & {\em  -31266.146407}\tabularnewline
Sc & $ \textrm{[Ar]} 4s^{2} 3d^{1} $ & -757.000629 & Pm & $ \textrm{[Xe]} 4f^{5} 6s^{2} $ & -9645.500832 & Md & $ \textrm{[Rn]} 5f^{14} 7s^{1} $ & -32014.176598\tabularnewline
Ti & $ \textrm{[Ar]} 4s^{2} 3d^{2} $ & -845.497930 & Sm & $ \textrm{[Xe]} 4f^{6} 6s^{2} $ & -10025.150892 & No & $ \textrm{[Rn]} 5f^{14} 7s^{2} $ & -32772.269829\tabularnewline
V & $ \textrm{[Ar]} 3d^{3} 4s^{2} $ & -939.796100 & Eu & $ \textrm{[Xe]} 4f^{7} 6s^{2} $ & -10413.476735 & Lr & $ \textrm{[Rn]} 5f^{14} 7s^{2} 6d^{1} $ & -33540.454380\tabularnewline
Cr & $ \textrm{[Ar]} 3d^{4} 4s^{2} $ & -1040.034946 & Gd & $ \textrm{[Xe]} 4f^{8} 6s^{2} $ & -10810.552897 & Rf & $ \textrm{[Rn]} 5f^{14} 6d^{2} 7s^{2} $ & -34318.854809\tabularnewline
Mn & $ \textrm{[Ar]} 4s^{1} 3d^{6} $ & -1146.366756 & Tb & $ \textrm{[Xe]} 4f^{9} 6s^{2} $ & -11216.453617 & Db & $ \textrm{[Rn]} 5f^{14} 6d^{4} 7s^{1} $ & -35107.525943\tabularnewline
Fe & $ \textrm{[Ar]} 4s^{1} 3d^{7} $ & -1258.917212 & Dy & $ \textrm{[Xe]} 4f^{10} 6s^{2} $ & -11631.252911 & Sg & $ \textrm{[Rn]} 5f^{14} 6d^{6} $ & -35906.506548\tabularnewline
Co & $ \textrm{[Ar]} 4s^{1} 3d^{8} $ & -1377.819755 & Ho & $ \textrm{[Xe]} 4f^{11} 6s^{2} $ & -12055.024619 & Bh & $ \textrm{[Rn]} 5f^{14} 6d^{7} $ & -36715.824635\tabularnewline
Ni & $ \textrm{[Ar]} 3d^{9} 4s^{1} $ & -1503.210775 & Er & $ \textrm{[Xe]} 4f^{12} 6s^{2} $ & -12487.842443 & Hs & $ \textrm{[Rn]} 5f^{14} 6d^{8} $ & -37535.505151\tabularnewline
Cu & $ \textrm{[Ar]} 3d^{10} 4s^{1} $ & -1635.226377 & Tm & $ \textrm{[Xe]} 4f^{13} 6s^{2} $ & -12929.779972 & Mt & $ \textrm{[Rn]} 5f^{14} 6d^{9} $ & -38365.584348\tabularnewline
Zn & $ \textrm{[Ar]} 3d^{10} 4s^{2} $ & -1773.909886 & Yb & $ \textrm{[Xe]} 4f^{14} 6s^{2} $ & -13380.910702 & Ds & $ \textrm{[Rn]} 5f^{14} 6d^{10} $ & -39206.098757\tabularnewline
Ga & $ \textrm{[Ar]} 3d^{10} 4s^{2} 4p^{1} $ & -1919.085911 & Lu & $ \textrm{[Xe]} 4f^{14} 6s^{2} 5d^{1} $ & -13840.976253 & Rg & $ \textrm{[Rn]} 5f^{14} 6d^{10} 7s^{1} $ & -40056.951158\tabularnewline
Ge & $ \textrm{[Ar]} 3d^{10} 4s^{2} 4p^{2} $ & -2070.946515 & Hf & $ \textrm{[Xe]} 4f^{14} 6s^{2} 5d^{2} $ & -14310.121254 & Cn & $ \textrm{[Rn]} 5f^{14} 6d^{10} 7s^{2} $ & -40918.195130\tabularnewline
As & $ \textrm{[Ar]} 3d^{10} 4s^{2} 4p^{3} $ & -2229.571620 & Ta & $ \textrm{[Xe]} 4f^{14} 6s^{2} 5d^{3} $ & -14788.392156 & Nh & $ \textrm{[Rn]} 5f^{14} 6d^{10} 7s^{2} 7p^{1} $ & -41789.700671\tabularnewline
Se & $ \textrm{[Ar]} 3d^{10} 4s^{2} 4p^{4} $ & -2395.043625 & W & $ \textrm{[Xe]} 4f^{14} 6s^{1} 5d^{5} $ & -15275.846800 & Fl & $ \textrm{[Rn]} 5f^{14} 6d^{10} 7s^{2} 7p^{2} $ & -42671.589032\tabularnewline
Br & $ \textrm{[Ar]} 3d^{10} 4s^{2} 4p^{5} $ & -2567.446685 & Re & $ \textrm{[Xe]} 4f^{14} 5d^{6} 6s^{1} $ & -15772.541265 & Mc & $ \textrm{[Rn]} 5f^{14} 6d^{10} 7s^{2} 7p^{3} $ & -43563.886976\tabularnewline
Kr & $ \textrm{[Ar]} 3d^{10} 4s^{2} 4p^{6} $ & -2746.866101 & Os & $ \textrm{[Xe]} 4f^{14} 5d^{8} $ & -16278.531177 & Lv & $ \textrm{[Rn]} 5f^{14} 6d^{10} 7s^{2} 7p^{4} $ & -44466.621119\tabularnewline
Rb & $ \textrm{[Kr]} 5s^{1} $ & -2932.972209 & Ir & $ \textrm{[Xe]} 4f^{14} 5d^{9} $ & -16793.845129 & Ts & $ \textrm{[Rn]} 5f^{14} 6d^{10} 7s^{2} 7p^{5} $ & -45379.818244\tabularnewline
Sr & $ \textrm{[Kr]} 5s^{2} $ & -3125.998090 & Pt & $ \textrm{[Xe]} 4f^{14} 5d^{10} $ & -17318.533845 & Og & $ \textrm{[Rn]} 5f^{14} 6d^{10} 7s^{2} 7p^{6} $ & -46303.505356\tabularnewline
Y & $ \textrm{[Kr]} 5s^{2} 4d^{1} $ & -3325.964742 & Au & $ \textrm{[Xe]} 4f^{14} 5d^{10} 6s^{1} $ & -17852.550237 &  & & \tabularnewline
Zr & $ \textrm{[Kr]} 5s^{2} 4d^{2} $ & -3533.076869 & Hg & $ \textrm{[Xe]} 4f^{14} 5d^{10} 6s^{2} $ & -18395.920112 &  & & \tabularnewline
\end{tabular}